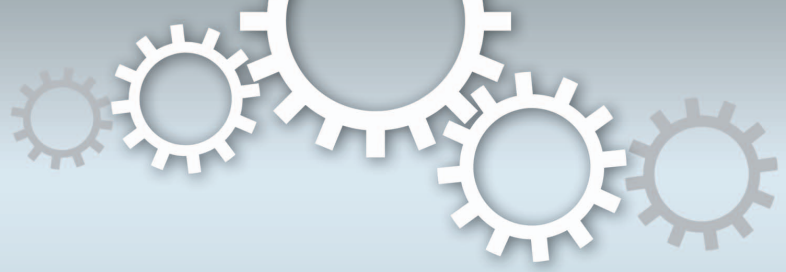

**OPEN**

# Novel Friction Law for the Static Friction Force based on Local Precursor Slipping

Yu Katano[1]*, Ken Nakano[2], Michio Otsuki[3] & Hiroshi Matsukawa[1]

[1]Department of Physics and Mathematics, Aoyama Gakuin University, 5-10-1 Fuchinobe, Sagamihara 252-5258, Japan, [2]Faculty of Environment and Information Sciences, Yokohama National University, 79-7 Tokiwadai, Hodogaya, Yokohama 240-8501, Japan, [3]Department of Materials Science, Shimane University, 1060 Nishikawatsu-cho, Matsue 690-8504, Japan.

Correspondence and requests for materials should be addressed to H.M. (matsu@phys. aoyama.ac.jp)

* Current address: Applied Technology Engineering Department 2, NetOneSystems Co., Ltd., JP TOWER, 2-7-2, Marunouchi, Chiyoda-ku, Tokyo 100-7024, Japan.

The sliding of a solid object on a solid substrate requires a shear force that is larger than the maximum static friction force. It is commonly believed that the maximum static friction force is proportional to the loading force and does not depend on the apparent contact area. The ratio of the maximum static friction force to the loading force is called the static friction coefficient $\mu_M$, which is considered to be a constant. Here, we conduct experiments demonstrating that the static friction force of a slider on a substrate follows a novel friction law under certain conditions. The magnitude of $\mu_M$ decreases as the loading force increases or as the apparent contact area decreases. This behavior is caused by the slip of local precursors before the onset of bulk sliding and is consistent with recent theory. The results of this study will develop novel methods for static friction control.

T he application of a shear force larger than the maximum static friction force to a slider on a base block causes the slider to start sliding. A kinetic friction force is exerted on the slider in the sliding state. Friction appears in various systems ranging from atomic to geological scales and has been studied since ancient times[1–6]. From an engineering perspective, friction control is critical to the effective functioning of machines and can enable advances in energy and global warming problems[4–6]. In the 15th century, da Vinci discovered that the friction force is proportional to the loading force and does not depend on the apparent contact area between two solid objects. About 200 years later, Amontons rediscovered these results, which are now collectively called Amontons' law. This law is commonly believed to hold well for diverse systems, discussed in high school textbook, and used to achieve friction control for various machines. The ratio of the maximum static friction force to the loading force is called the static friction coefficient. When Amontons' law holds, this ratio is a constant that does not depend on the loading force or the apparent contact area.

Friction mechanisms are usually explained as follows[1–6]. The roughness of the solid surfaces in contact with each other results in a very small total area for the real contact points relative to the apparent contact area. The real contact points yield a finite shear strength at the interface, which is produced by interatomic or intermolecular forces. Relative sliding motion between two solids in contact requires the application of a shear force that is larger than the total shear strength, that is, the maximum static friction force must be applied. The total area of the real contact points is called the real contact area. It is considered that the real contact area is proportional to the loading force and does not depend on the apparent contact area, and the shear strength per unit of the real contact area is constant. Thus, the friction force is proportional to the loading force, does not depend on the apparent contact area, and Amontons' law holds. Note that, to date, alternative schemes have also been put forward to explain the mechanisms of friction and Amontons' law[7].

Recently, the contact interface has become an interesting topic in the case of friction for a shear force that is smaller than the macroscopic maximum static friction force $F_s^{max}$, which corresponds to the onset of sliding motion for the whole slider[8–12]. Bulk sliding does not occur under this condition. The instantaneous and local densities of the real contact area have been measured by employing the transmission and reflection of a laser sheet, demonstrating that local precursor slipping occurs at the contact interface between a polymethyl-methacrylate (PMMA) slider and a PMMA base block[8–12]. A shear force is applied to the slider by a pushing rod or a spring at the trailing edge of the slider. The precursor front begins from the trailing edge of the slider and stops after propagating a certain distance. The front of the next precursor also begins from the trailing edge after a certain interval and stops after a propagation length that is longer than the previous length. Initially, the propagation length gradually increases with the number of precursors; however, after exceeding a critical length, the propagation length begins to grow rapidly. When the precursor front reaches the leading edge of the slider, bulk sliding occurs. This behavior has been observed in various experiments[8–12], and the front propagation velocity has been investigated in experiments[8–11] and in numerical and theoretical studies[12–19].

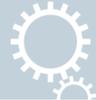

Ben-David *et al.* reported experimentally that the macroscopic static friction coefficient $\mu_M = F_s^{max}/W$, where $W$ is the loading force, are strongly dependent on precursor dynamics[11]. On the other hand, finite element method (FEM) approaches and analytical calculations have been used to show that $\mu_M$ decreases as the pressure and the system length $L$ increase[19]. These behaviors are determined by the precursor dynamics, and the quasi-static precursor transitions into rapid motion at a critical length $l_c$. In the calculations, Amontons' law is assumed to hold locally, i.e., the local frictional stress is proportional to the local pressure. The following relation was obtained,

$$\mu_M = \mu_k + (\mu_s - \mu_k)\ l_c/L, \tag{1}$$

when $l_c/L$ is smaller than unity[19]. Here, $\mu_s$ and $\mu_k$ are the local static and kinetic friction coefficients, respectively. The dependence of $l_c$ on the loading force yields a new friction law:

$$\mu_M = \mu_k + aW^{-1/3}, \tag{2}$$

where $a$ is a constant that does not depend on $W$. When the magnitude of $W$ and $L$ is such that $l_c/L$ is close to 0 or unity, Amontons' law holds approximately. The aforementioned relationships have been obtained from an analytical treatment of the 1D effective model and have been verified by a 2D FEM calculation. However, these results have not yet been verified experimentally.

In this study, we experimentally investigate the precursor dynamics and the macroscopic static friction coefficient of a PMMA slider loaded on a PMMA base block using the transmission of a laser sheet. The macroscopic static friction coefficient is found to decrease as the loading force increases. The experimental results show little dependence on the driving velocity or the height from the contact interface at which the tangential spring force, which we call the shear force, is applied. The results verify relationships (1) and (2) within the experimental accuracy, that is, the breakdown of Amontons' law and the validity of the novel friction law, Eq. (2). It is also indicated that how these relationships are associated with the local precursor slip. The results obtained herein are consistent with previous experimental studies[20] including the work by Ben-David *et al.* noted above[11]. Moreover, the macroscopic static friction coefficient decreases as the apparent contact area decreases. This result also indicates the breakdown of Amontons' law.

## Results

We apply a shear force to the slider by a leaf spring at a height $h$ from the contact surface with a constant driving velocity $V$, and a uniform normal force $F_z$. Figure 1a shows the typical time evolution of the shear force for $V = 0.4$ mm/s, $h = 10$ mm, and $F_z = 400$ N. After the leaf spring starts moving, the shear force increases linearly from 0 over time $t$, which means that the shear force balances the static friction force; the slider is at rest in this regime. The shear force drops immediately after it reaches its maximum value at $t = 5.8$ s. The drop is caused by the sliding motion of the whole slider and is followed by the arrest of the slider. The maximum value of the shear force is the macroscopic static friction force $F_s^{max}$. The shear force increases linearly and periodically drops sharply, that is, a periodic stick-slip motion appears. The inset of Fig. 1a displays an enlarged view of $F_x$ near $t = 5$ s. Prior to the large drop in the shear force corresponding to the onset of bulk sliding, there is a sequence of small drops, which are shown by black arrows. This behavior indicates that the precursors appear as partial slipping when the shear force is smaller than the macroscopic maximum static friction force.

The spatio-temporal distribution of the real contact area density in Fig. 1b clearly shows the sequence of discrete precursor slips that start from the trailing edge of the slider. This behavior results from the relatively small pressure at the bottom of the trailing edge caused by the torque induced by the shear force, even though the applied pressure is uniform[17–19]. Each precursor front stops after propagating a certain length, as indicated by the red circles in the figure. The propagation length of the subsequent precursor exceeds that of the preceding precursor. When the precursor front reaches the leading edge of the slider, bulk sliding occurs, and the shear force drops dramatically.

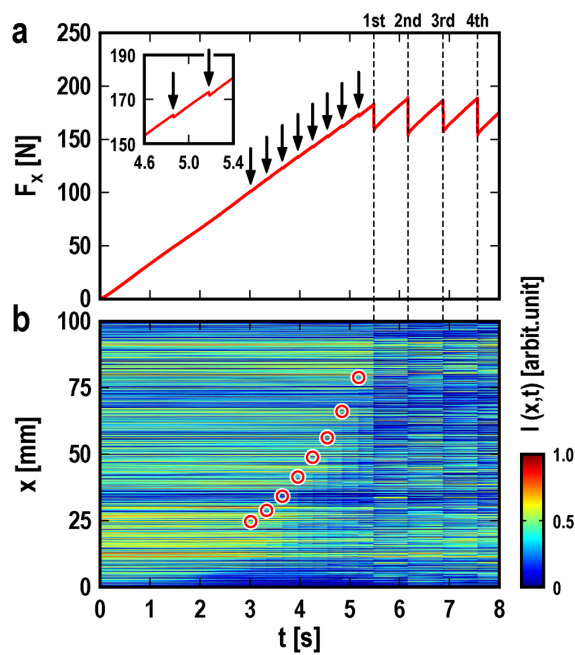

**Figure 1 | Time evolution of the shear force and the instantaneous and local densities of the real contact area.** (a) The shear force and (b) the instantaneous and local densities of the real contact area are shown for $V = 0.4$ mm/s, $h = 10$ mm, and $F_z = 400$ N. The inset of (a) displays an enlarged view of $F_x$ near $t = 5$ s. The arrows indicate when local precursor slipping occurs, the vertical broken lines show when bulk sliding occurs, and the red circles denote where local precursor slipping stops.

Figure 2 shows the propagation length $l$ of the precursors normalized by the slider length $L$ as a function of the ratio of the shear force $F_x$ to the loading force $W$ for various magnitudes of the driving velocity $V$ (Fig. 2a), various heights $h$ at which the shear force is applied (Fig. 2b), and various applied normal forces $F_z$ (Fig. 2c). The loading force $W$ is the sum of the applied normal force $F_z$ at the top plane of the slider and the weight of the slider $Mg = 6$ N, where $M$ is the mass of the slider and $g$ is the acceleration of gravity. Initially, $l$ increases linearly with small slope with increasing $F_x/W$; however, $l$ grows rapidly beyond a certain value of $F_x/W$ for every condition. The two solid straight lines show the results of fitting the data in the ranges of small and large values of $F_x/W$, respectively, for all conditions in (a) and (b). In (c) the two solid straight lines in the same color show the results of the fitting the data in the small and large values of $F_x/W$, respectively, for every magnitude of $F_z$. We can define the critical length normalized by the system length $l_c/L$ from the intersection of the two solid straight lines. The magnitudes of $l_c/L$ are indicated by broken horizontal lines in the figures and correspond to the critical length of the slow increase of the propagation length of the precursor front. The magnitude of $l_c/L$ does not depend on the driving velocity or the height $h$ and is approximately 0.45 for $F_z = 400$ N. When the precursor front reaches the leading edge of the slider, bulk sliding occurs. The value of $F_x/W$ at $l/L = 1$ corresponds to the macroscopic static friction coefficient $\mu_M$. Figure 2c shows $l/L$ as a function of $F_x/W$ for various normal forces. The critical length $l_c$ appears for any value of the applied normal force, but the magnitude

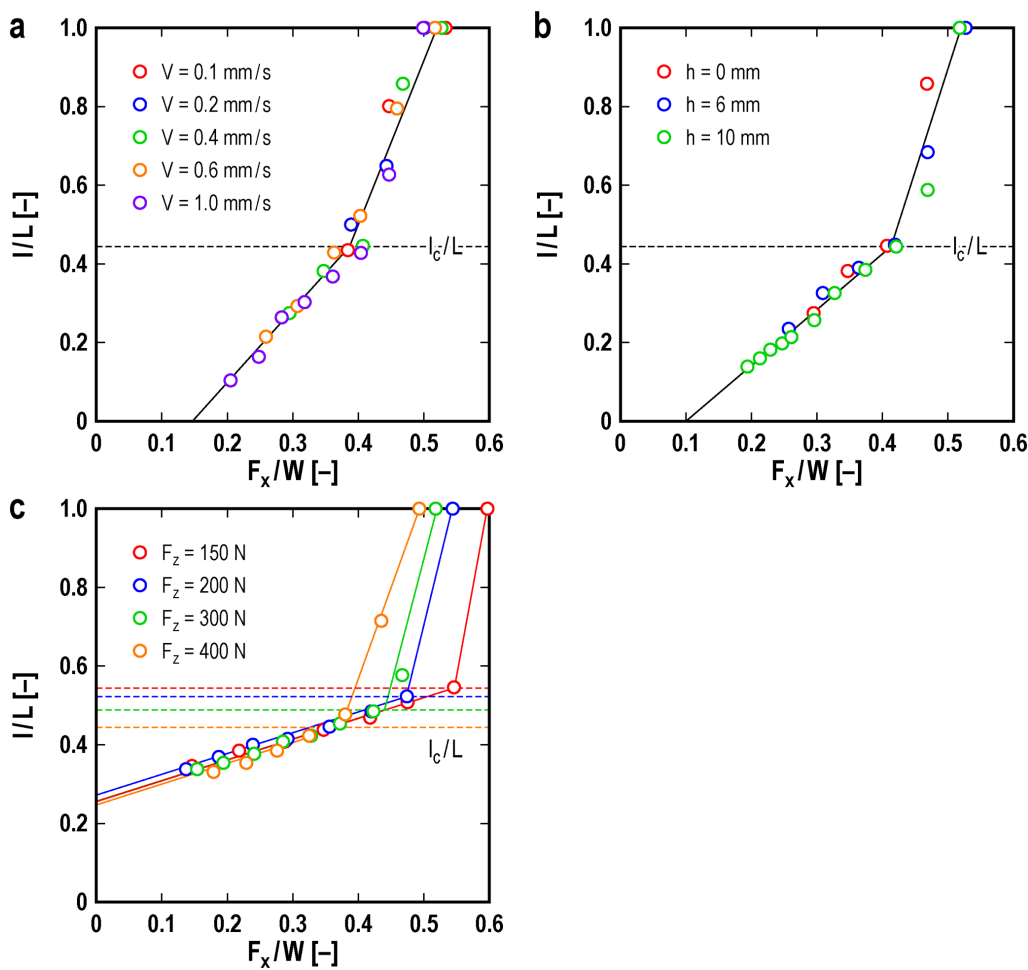

**Figure 2 | Propagation length $l$ vs. shear force $F_x$.** The propagation length $l$ of the precursors, normalized by the slider length $L$, is shown as a function of the shear force $F_x$, which is normalized by the loading force $W$, for various magnitudes of (a) the driving velocity $V$, (b) the height $h$ at which the shear force is applied, and (c) the applied normal force $F_z$. In (a), $F_z = 400$ N and $h = 0$ mm; in (b), $F_z = 400$ N and $V = 0.4$ mm/s; and in (c), $V = 0.4$ mm/s and $h = 0$ mm. The two solid straight lines show the results of fitting the data in the ranges of small and large values of $F_x/W$, respectively, for all conditions in (a) and (b) because the behaviors do not depend on $V$ and $h$. In (c) the two solid straight lines in the same color show the results of the fitting the data in the small and large values of $F_x/W$, respectively, for every magnitude of $F_z$ because the behaviors depends on $F_z$. The broken lines show the critical length $l_c$ normalized by $L$ determined by the intersection of the two solid straight lines.

of $l_c$ depends on the applied normal force. This result will be discussed in more detail below.

We also measured the magnitude of the macroscopic static friction coefficient $\mu_M$, which corresponds to the onset of bulk sliding of the slider, under various conditions. Figure 3 shows the dependence of $\mu_M$ on the loading force for various values of the driving velocity $V$ (Fig. 3a) and height $h$ (Fig. 3b). Each point in the figure represents the average value of 3 to 9 (typically 9) measurements. Amontons' law states that $\mu_M$ does not depend on the loading force. However, Fig. 3 shows that $\mu_M$ decreases as the loading force increases for all values of $V$ and $h$. This behavior indicates the breakdown of Amontons' law and is consistent with the results obtained from the FEM and analytical calculations[19]. The driving velocity $V$ and height $h$ have little effect on the magnitude of $\mu_M$. Note that the apparent contact area between the slider and the base block changes negligibly over the range of loading forces used in the experiment.

The dependence of $\mu_M$ on the loading force is well-fitted by the following formula:

$$\mu_M = aW^n + b. \tag{3}$$

Here, $a$, $b$, and $n$ are fitting parameters. The broken lines in Fig. 3 show the fitting results for $V = 0.4$ mm/s and $h = 0$ mm. The parameter values obtained by employing the weighted least squares method are $a = 0.53 \pm 0.15$, $b = 0.32 \pm 0.24$, and $n = -0.16 \pm 0.15$, which exhibit little dependence on the experimental conditions. The magnitude of the errors corresponds to the standard deviation. Reference 19 reports $n = -1/3$, as noted in Eq. (2), obtained from an analytical method based on the 1D effective model. The result is verified by a 2D FEM calculation. The aforementioned experimentally obtained value of $n$ is quantitatively consistent with the theoretical result within the experimental accuracy. Thus, the novel friction law given by Eq. (2) is experimentally verified.

Reference 19 also predicts a relationship between $\mu_M$ and $l_c$, Eq. (1). Figure 4 shows the experimental results for $\mu_M$ as a function of $l_c/L$. The figure shows that $\mu_M$ is determined by $l_c/L$, which means that the dependence of $\mu_M$ on the loading force is caused by the precursor slip. In fact, $l_c/L$ decreases with increasing the loading force as shown by the inset of Fig. 4, which causes the decrease of $\mu_M$ with increasing the loading force. The linear relation between $\mu_M$ and $l_c$ shown in the figure is consistent with the theoretical prediction, Eq. (1)[19]. We used the least squares method to fit the following formula to the experimental data:

$$\mu_M = \alpha l_c/L + \beta. \tag{4}$$

Here, $\alpha$ and $\beta$ are fitting parameters. The broken line in Fig. 4 shows the fitting results for the parameters, $\alpha = 0.89 \pm 0.15$ and $\beta = 0.092$ $\pm 0.077$. Equation (4) fits the experimental data fairly well. In ref. 19, both $b$ in Eq. (3) and $\beta$ in Eq. (4) correspond to the local kinetic friction coefficient $\mu_k$. Therefore, we can estimate the value of $\mu_k$ by employing two independent methods. The two estimated values are equal to each other within the experimental accuracy. Thus, the theoretical predictions have been verified. The value of $\alpha$ is predicted to correspond to $\mu_s - \mu_k$, as shown in Eq. (2)[19]. The values of $\mu_s$ and $\mu_k$ obtained from fitting Eq. (4) to the experimental data are $\mu_s = 0.98 \pm 0.17$ and $\mu_k = 0.092 \pm 0.077$, which are consistent with a previous experimental study[11].

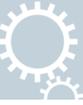

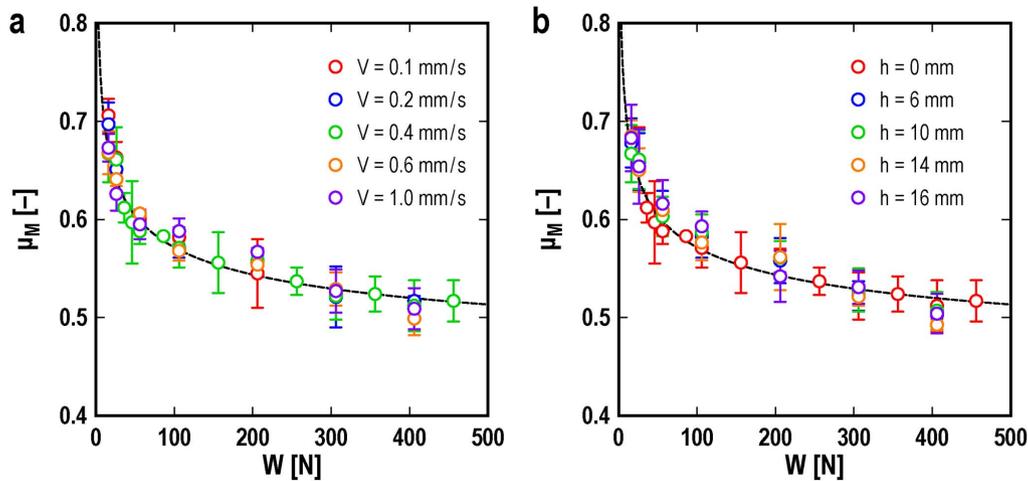

**Figure 3 | Macroscopic static friction coefficient $\mu_M$ vs. loading force $W$.** The dependence of the macroscopic static friction coefficient $\mu_M$ on the loading force $W$ is shown for various values of (a) the driving velocity $V$ and (b) the height $h$. The error bars indicate the standard deviation. In (a), $h = 0$ mm, and in (b), $V = 0.4$ mm/s. The broken lines show the results obtained by fitting the data with Eq. (3) for $V = 0.4$ mm/s and $h = 0$ mm.

Amontons' law also states that the friction coefficient does not depend on the apparent contact area. We investigate the dependence of the apparent contact area of the macroscopic static friction coefficient. Figure 5 shows the $\mu_M$ values plotted as a function of various lengths of the sliding surface $L'$, i.e., for various magnitudes of the apparent contact area. To perform this experiment, two regions of the lower portion of the slider are removed. The shape of the slider is shown in the Supplemental Methods. The figure clearly shows that the macroscopic static friction coefficient depends on the apparent contact area. This result also indicates the breakdown of Amontons' law.

## Discussion

In this study, we investigate the frictional behavior of a solid slider on a solid base block. We observe precursor slipping before the onset of bulk sliding; each of these slips causes a small drop in the shear force applied to the slider. The front of the precursor slip starts from the trailing edge of the slider and stops after propagating a length $l$. After a certain interval, the next precursor starts to slip from the trailing edge and stops after a propagation length $l$, which is longer than the propagation length of the previous precursor. Initially, the magnitude of the propagation length $l$ increases gradually; however, after exceeding a critical length $l_c$, the propagation length increases rapidly. The magnitude of $l_c$ depends on the loading force $W$. We also find that the macroscopic static friction coefficient $\mu_M$ decreases as $W$ increases. The decrease of $\mu_M$ with increasing $W$ is consistent with the data of the recent work by Ben-David et al[11], although they did not discuss this behavior. This behavior indicates the breakdown of Amontons' law and is consistent with 2D FEM and analytical calculations[19]. Reference 19 predicts a novel friction law, Eq. (2), which implies that the macroscopic static friction coefficient is the sum of a term that is proportional to $W^{-1/3}$ and the local kinetic friction coefficient $\mu_k$. The $-1/3$ power of the $W$ dependence of $\mu_M$ is consistent with the present experimental results within the experimental accuracy. Hence, the novel friction law, Eq. (2), is experimentally verified.

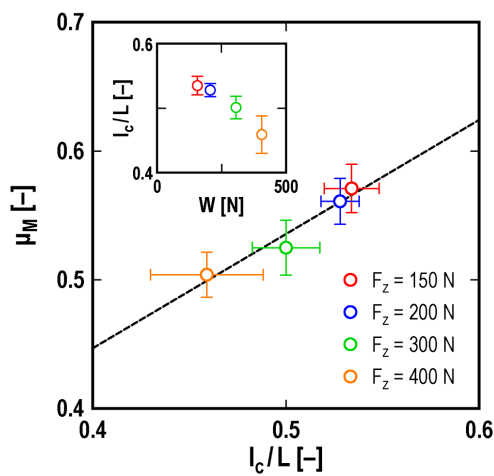

**Figure 4 | Macroscopic static friction coefficient $\mu_M$ vs. critical precursor length $l_c$.** The dependence of the macroscopic static friction coefficient $\mu_M$ is shown for the critical propagation length of the precursor $l_c$, which is normalized by the slider length $L$. The broken line shows the fitting curve obtained by Eq. (4). The inset shows the dependence of $l_c/L$ on the loading force $W$. The experimental parameters are $V = 0.4$ mm/s and $h = 0$ mm.

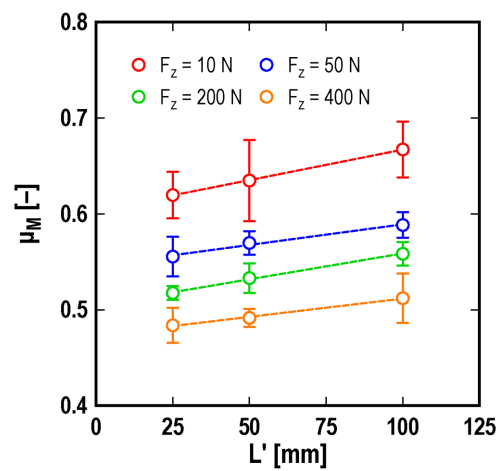

**Figure 5 | Macroscopic static friction coefficient $\mu_M$ vs. sliding surface length $L'$.** The dependence of the macroscopic static friction coefficient $\mu_M$ is shown for the length of the sliding surface $L'$, i.e., the apparent contact area. The length of the upper surface of the slider is held constant at $L = 100$ mm. The error bars indicate the standard deviation. The experimental parameters are $V = 0.4$ mm/s and $h = 0$ mm.

The linear relation between $\mu_M$ and $l_c$ is also consistent with the theoretical prediction[19] and indicates that the dependence of $\mu_M$ on the loading force is caused by the precursor slip. Two independent methods yield the same estimate of the local kinetic friction coefficient $\mu_k$ within the experimental accuracy and support the theoretical prediction. The magnitude of $\mu_M$ depends on the apparent contact area, which also indicates the breakdown of Amontons' law.

In the 2D FEM and 1D theoretical calculations[19], the quasi-static precursor slip appears and grows continuously from the trailing edge of the slider. When the front of the quasi-static precursor reaches a critical length $l_c$, the quasi-static precursor becomes unstable and changes to a leading rapid precursor, which starts to move at a high velocity toward the leading edge of the slider. When the front of the precursor reaches the leading edge, bulk sliding occurs. In ref. 19, Amontons' law is assumed to hold locally, and the local friction coefficient is considered to be a function of the instantaneous local velocity only. When the local velocity becomes finite, the magnitude of the local friction coefficient decreases linearly with the local slip velocity from the value of the local static friction coefficient $\mu_s$. Above a certain velocity, the local friction coefficient takes a constant value that is equal to the local kinetic friction coefficient $\mu_k$. After the high-velocity bulk sliding occurs, the magnitude of the ratio of the local shear stress to the local pressure is maintained at the value corresponding to the high-velocity bulk sliding due to the finite relaxation time caused by the viscosity; this ratio coincides with $\mu_k$. However, the magnitude of the ratio of the local shear stress to the local pressure coincides with $\mu_s$ in the region through which the quasi-static precursor has passed because the slip velocity and the slip distance of the precursor are sufficiently small. The value of $\mu_M$ is given by the distribution of the shear stress that occurs when the front of the quasi-static precursor reaches the critical value $l_c$. Thus, $\mu_M$ is determined by $l_c$, and the linear relation between these two variables, Eq. (1), holds. The dependence of $l_c$ on the loading force $W$ yields the dependence of $\mu_M$ on $W$. The magnitude of $l_c$ is determined by competition between the stabilization and destabilization of the quasi-static precursor. The stabilization is caused by the viscosity, and the destabilization is caused by the local frictional stress, which reduces with the velocity. The destabilization effect increases with $W$. A dimensional analysis on the primary terms of the time evolution of the fluctuation around the quasi-static precursor solution shows that $l_c/L$ is proportional to $W^{-1/3}$. Thus, the expression for $\mu_M$ contains a term that is proportional to $W^{-1/3}$. When the magnitude of $W$ and $L$ is such that $l_c/L$ is close to 0 or unity, Amontons' law holds approximately.

The effect of the precursor dynamics on $\mu_M$ and the dependence of $\mu_M$ on the size of the slider and loading conditions have been discussed in ref. 21. The model employed in that work, however, does not take into account the effect of torque induced by the shear force and then the non-uniformity of the pressure at the bottom of the slider, which causes the transition of the precursor slip at $l_c$ in ref. 19 and in the present work as discussed previously. The mechanism of the change of static friction coefficient in ref. 21 is considered to be different from that in ref. 19 and in the present work.

In experiments, including the present study[9–12], however, individual precursor events are observed to occur discretely, while a quasi-static precursor has not been observed. These results may be attributed to the finite resolution of the experiments and the small slip distance of the quasi-static precursor. These factors could prevent the observation of a quasi-static precursor in the experiments. Thus, a precursor with a propagation length below the critical length observed in the experiments would correspond to a bounded rapid precursor, which has been observed in 2D FEM and 1D theoretical calculations, and has a finite propagation velocity, but stops beyond a certain length[19]. The transition of the bounded rapid precursor to the leading rapid precursor occurs near $l_c$. Other possible mechanisms for the disappearance of the quasi-static precursor in experiments are the discrete nature of each real contact point or the slight deviation of the local friction law assumed in the calculations[19] from the actual local friction law. As previously mentioned, the local frictional stress is determined by the local pressure and the instantaneous slip velocity in ref. 19. In the actual system, the friction force exhibits hysteresis[22,23]. These effects could transform the quasi-static continuous precursor into a discrete precursor. As previously mentioned, the dependence of $l_c$ on the loading force $W$ results from the dimensional analysis, and the same $W$ dependence of $l_c$ is expected to hold even in the case of the disappearance of the quasi-static precursor.

In the experiments[9–12], a precursor with a propagation length that is larger than the critical length can stop in the slider. This behavior has not been predicted in the FEM calculations[19]. However, FEM calculations were performed over a limited range of slider lengths, and calculations for longer sliders may show behavior similar to that observed in the experiments. After a precursor with a propagation length exceeding the critical length $l_c$ stops, it should leave an imprint of the transition at $l_c$. The experimentally observed linear relationship between the macroscopic static friction coefficient and the critical length $l_c$ indicates that the imprint corresponds to the stress distribution when the propagation length of the precursor reaches $l_c$, which is maintained after the subsequent precursor has passed. Thus, the agreement between the experimental results obtained here and the calculation still holds.

In conclusion, we have shown that the macroscopic static friction coefficient of solid objects decreases as the loading force increases and as the apparent contact area decreases. These results indicate the breakdown of Amontons' law and the validity of a novel friction law, Eq. (2). This behavior is caused by precursor slipping occurring prior to bulk sliding. The observed results are consistent with the predictions of 2D FEM and 1D analytical calculations[19]. The results obtained in this study can lead to new methods for controlling the maximum static friction force.

## Methods

The transmission of a laser sheet is used to measure the instantaneous and local densities of the real contact area between a slider and a base block, which are both composed of PMMA. The linear contact area between the slider and the base block is 100 mm long and 0.8 mm wide. A normal force $F_z$ is applied uniformly at the top surface of the slider. The apparatus and experimental methods applied herein are similar to those used in ref. 12, except for the driving method. The slider is pushed by a leaf spring via a stainless L-shaped arm at a height $h$ from the contact surface. The leaf spring, with a spring constant $k = 95$ kN/m, is moved with a constant driving velocity $V$ ranging from 0.1 to 1.0 mm/s. Details of the experimental apparatus and methods are given in the Supplemental Methods.

## Acknowledgments

The authors thank to S. Maegawa, N. Kado, and C. Tadokoro for valuable discussions and technical supports. This work was financially supported by KAKENHI (22540398), (25800220), and (26400403) from MEXT.

## Author contributions

Y.K. performed the experiments. Y.K. and K.N. prepared all figures. Y.K., K.N., M.O. and H.M. contributed to the analysis, discussions, and preparation of the manuscript.

## Additional information

**Supplementary information** accompanies this paper at http://www.nature.com/scientificreports

**Competing financial interests:** The authors declare no competing financial interests.

**How to cite this article:** Katano, Y., Nakano, K., Otsuki, M. & Matsukawa, H. Novel Friction Law for the Static Friction Force based on Local Precursor Slipping. *Sci. Rep.* **4**, 6324; DOI:10.1038/srep06324 (2014).

Novel Friction Law for the Static Friction Force based on Local Precursor Slipping

Yu Katano[1+], Ken Nakano[2], Michio Otsuki[3] & Hiroshi Matsukawa[1]*

[1]Department of Physics and Mathematics, Aoyama Gakuin University, 5-10-1 Fuchinobe, Sagamihara 252-5258, Japan

[2]Faculty of Environment and Information Sciences, Yokohama National University, 79-7 Tokiwadai, Hodogaya, Yokohama 240-8501, Japan

[3]Department of Materials Science, Shimane University, 1060 Nishikawatsu-cho, Matsue 690-8504, Japan

[+]Present address: Applied Technology Engineering Department 2, NetOneSystems Co., Ltd., JP TOWER, 2-7-2, Marunouchi, Chiyoda-ku, Tokyo 100-7024, Japan

* Correspondence and requests for materials should be addressed to H.M.

(matsu@phys.aoyama.ac.jp)

Supplemental Methods

A schematic of the apparatus is shown in Supplementary Fig. 1. The slider and base block are transparent rectangles composed of PMMA with a Young's modulus of 2.5 GPa. The length ($x$), width ($y$), and height ($z$) of the slider are 100, 10, and 20 mm, respectively, and the dimensions of the base block are 120 mm ($x$), 30 mm ($y$), and 40 mm ($z$). The slider is placed on the base block, which is fixed at the center of the apparatus. The bottom edge of this slider has an apex angle of 158°, and the contact area between the slider and the base block is linear in the $x$ direction. The linear contact area between the slider and the base block is 100 mm long and 0.8 mm wide. A normal force was applied to the slider via an elastic block composed of silicon rubber with a length, width, and height of 100, 10, and 10 mm, respectively. This elastic block is used to apply a uniform normal force to the top of the slider. Grease is also smeared on the contact surface between the slider and the elastic block to facilitate smooth sliding between the elastic block and the slider. Two load cells separated by 60 mm are used to measure the normal force. The observed values of each load cell are unchanged with 1% of precision, even after bulk sliding of the slider. A shear force is applied to the slider by a leaf spring, with a spring constant $k = 95$ kN/m, via a stainless L-shaped arm, which is fixed to the slider with glue at a height $h$ from the top surface of the base block, as shown in Supplementary Fig. 2. The shear force is measured by a strain gauge attached to the leaf spring.

The contact surfaces of the slider and the base block are polished in the $y$ direction using sandpaper (#240). The specimens are then washed using detergent, water, and ethanol and are allowed to dry naturally.

The apparatus has a transmissive optical system. The contact area is illuminated by a collimated laser sheet from a green laser diode with a wavelength $\lambda = 532$ nm. A high-speed camera measures the intensity of the transmissive light. The incident angle of the laser sheet to the contact zone is 60°, which is greater than the critical angle for total reflection at a PMMA-air interface (42°). Therefore, the incident light illuminating the real contact points is transmitted through the interface, and the incident light illuminating the regime that does not contain the real contact points is reflected at the interface. The measured intensity of the transmissive light can be used to quantitatively estimate the local real contact area density. The linear relationship between the increment in the total intensity of the transmissive light and the magnitude of the applied normal load is verified in the absence of a shear force.

The sampling frequency of the data logger is 500 Hz, and all of the measurements are performed in an air-conditioned room with a temperature and relative humidity of approximately 25°C and 30–40%, respectively.

The apparent contact area dependence of the macroscopic static friction coefficient is examined using the slider, from which regions of the lower portion are removed, as shown in Supplementary Fig. 3. The length of the upper surface of the slider is held constant at 100 mm.

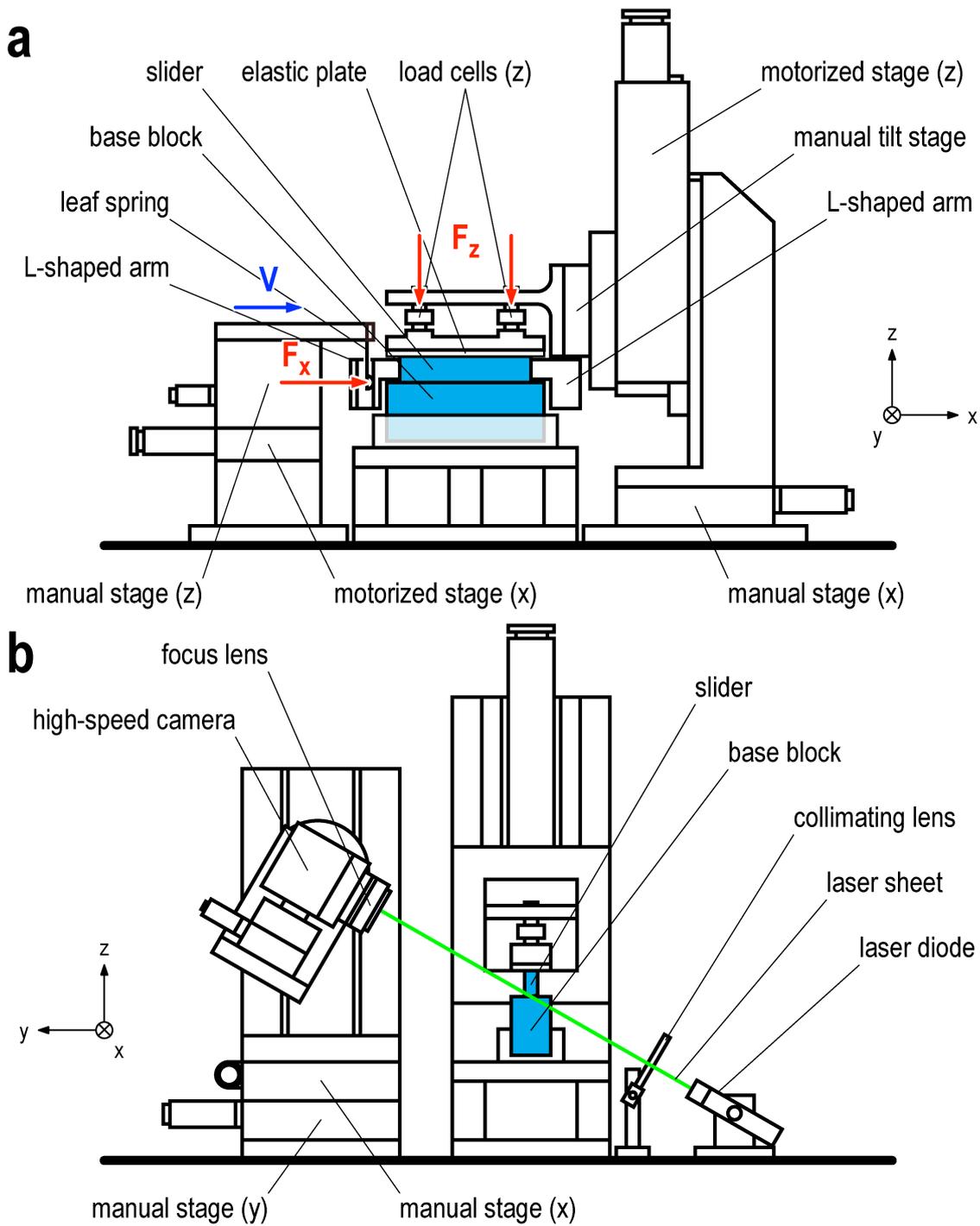

**Supplementary Fig. 1: Schematic of the experimental apparatus: (a) front view and (b) side view.**

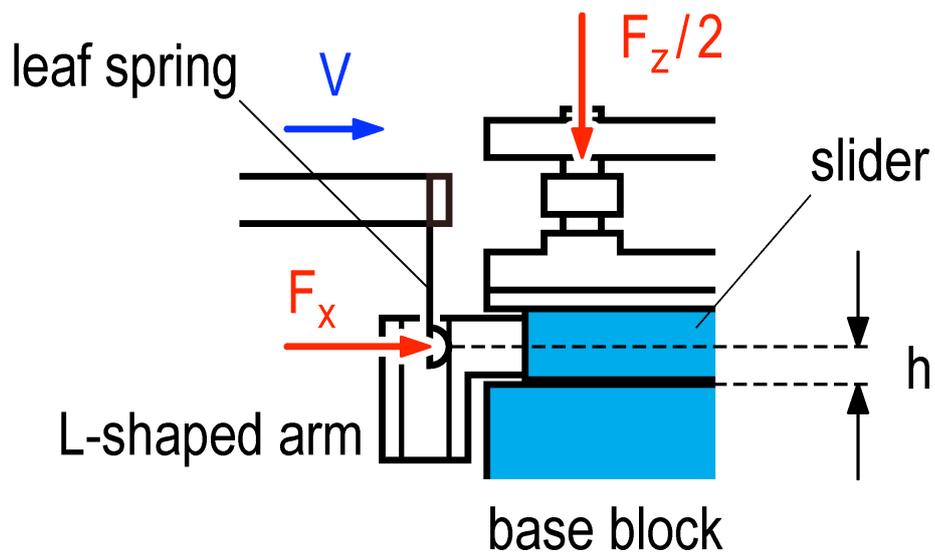

**Supplementary Fig. 2: Magnified view of the driving component.**

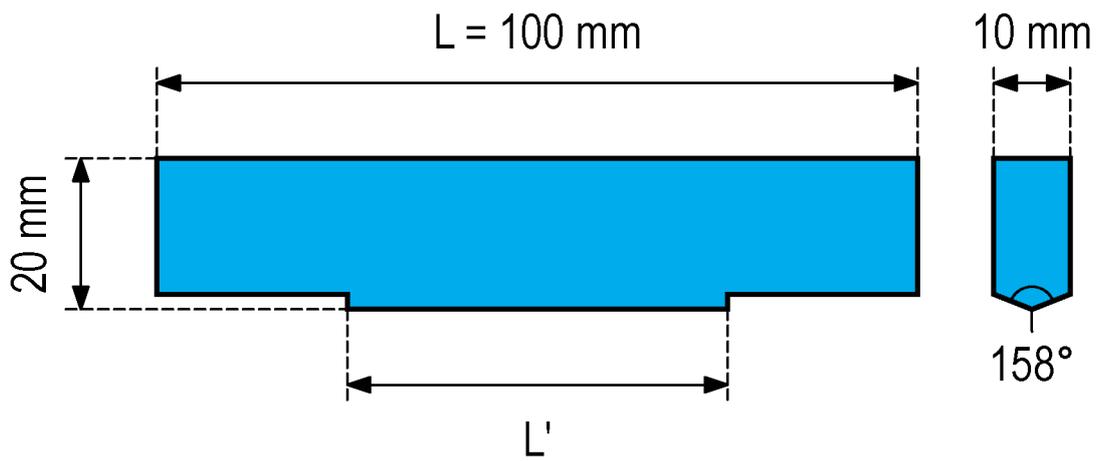

**Supplementary Fig. 3: Shape of slider used to investigate the apparent contact area dependence of the static friction coefficient.**